\documentclass[12pt]{iopart}

\usepackage[utf8]{inputenc}
\usepackage{xcolor}

\usepackage{multirow}
\usepackage{graphicx}
\usepackage{subcaption}

\usepackage{color,hyperref}

\begin{document}

\title[An Electroacoustic Lumped Element Model of a Dielectric Elastomer Membrane]{An Electroacoustic Lumped Element Model of a Dielectric Elastomer Membrane}

\author{C. Solano}
\address{Department of Mechanical Engineering, FAMU-FSU College of Engineering, Tallahassee, FL 32310}
\author{Y. Zhang and L. N. Cattafesta III}
\address{Mechanical, Materials, and Aerospace Engineering Department, Illinois Institute of Technology, Chicago, IL 60616 }
\ead{cad12e@fsu.edu}
\vspace{10pt}

\begin{abstract}
Dielectric elastomers are widely studied for their use in robotic and medical devices due to their shape changing properties. Recently, they have also been incorporated into acoustic devices, motivating the development of an electroacoustic model for dielectric elastomers. This paper provides a lumped element model based on a prestretched membrane approximation of a dielectric elastomer actuated by a dc voltage. The electroacoustic model is validated via experiments using a laser Doppler vibrometer and acoustic impedance tube measurements. Good agreement between the experiments and model are demonstrated. The resulting validated model is expected to be useful in design optimization for acoustic applications.

\end{abstract}

%
%
%
%
%


\section{Introduction}
Acoustic liners are a type of noise absorber commonly used in aircraft engines. Relatively simple acoustic liners are comprised of a perforate face sheet (FS) backed by honeycomb cells and terminated with a rigid backing. These single degree of freedom (SDOF) devices have an effective noise suppression bandwidth of approximately one octave \cite{hubbard1995Noise}. Absorption can be increased at the expense of added size and weight to approximately two octaves by increasing the honeycomb depth and adding a second perforate layer, creating what is known as a two degree of freedom (2DOF) liner \cite{hubbard1995Noise}. More complex liners have been formulated to further increase absorption bandwidth. For example, active liners have been created that can change their physical dimensions, such as modifying the cavity volume \cite{Neise_1980, mcdonald1992method, MATSUHISA_1992} or changing the facesheet hole area \cite{Gaeta_Jr__1998, Nagaya_2001}. These types of liners are able to change their characteristic resonance frequency. Another well known type of adaptive liner is the bias flow liner. In this design, flow enters from the back of the cavity and exits through the FS resulting in a wide tunable sound absorption range as a function of the variable flow rate \cite{dean1975duct, Jing_1999}. However, active designs that require moving parts or bias flow can be heavy, complex, and require energy. 

With the advancement of smart materials, researchers have devised creative ways to simplify adaptive liner technology while still obtaining broadband absorption. For example, changes in liner geometry have been performed through the use of shape memory materials. The cavity volume was varied using shape memory polymers in Hermiller et al. \cite{Hermiller_2013}, with a resultant 500 - 600 Hz modification of the resonance frequency. Kreitzman et al. \cite{Kreitzman_2020} and Dodge et al. \cite{Dodge_2022} incorporated a shape memory alloy wire into a multilayered FS, changing the effective hole diameter of the FS and modifying the resonance frequency by 350 Hz. Liu et al. \cite{Liu_2007} replaced the rigid backing of a simplified acoustic liner (a Helmholtz resonator) with a piezoceramic diaphragm coupled to a passive electrical shunt network, allowing a tunable absorption range. Although compact and simple, a major disadvantage is that the piezoceramic disc is stiff relative to the cavity. This provides poor coupling, which limits the tunable range of the device. Their results motivated a softer active material that would better couple with the acoustics of the cavity. An example of a compliant material with modifiable properties is a dielectric elastomer (DE) - a smart material capable of changing shape when subjected to an electric field \cite{Pelrine_1998, Pelrine_2000, Kornbluh, Kornbluh_1999}. This material has been incorporated into acoustic liners by Abbad et al. \cite{Abbad_2018} and Dodge et al. \cite{Dodge_2021}. Abbad et al. \cite{Abbad_2018} replaced the rigid FS with a DE and was able to actuate the DE to modify the solid portion of the FS compliance. This shifted the resonance of the liner by 32 Hz and also achieved sound attenuation below 500 Hz. The work in Dodge et al. \cite{Dodge_2021} split the cavity of an acoustic liner with a DE and saw a shift in the resonance frequency of approximtately 100 Hz, or 11\%.

Researchers have incorporated the stress reduction of a DE into high fidelity finite element models to simulate the hyperelastic properties of dielectric elastomers \cite{wissler2007modeling, Park_2012} when subjected to electric fields. Analytical models encompassing the complexities of hyperelastic material properties of DEs were developed using a Kelvin model \cite{Sarban_2012} and Kelvin Voigt model \cite{Hoffstadt_2015} to capture its response to an electric field in terms of speed and relaxation \cite{Rizzello_2020, Kiser_2016}. Others have developed a lumped parameter model for strip-shaped dielectric elastomer membrane transducers  \cite{Rizzello_2020}. Further simplified models have incorporated variations of the stress expression, Eq. \ref{DE stress}, into dynamical models to determine how the resonance frequency of a DE is affected by voltage \cite{Dubois_2008}. These models are concerned with the mechanical response of the DE when subjected to a variable voltage loading and how its viscoelastic properties affect the corresponding time response in terms of speed and relaxation. 

The current paper is specifically focused on developing an electroacoustic lumped element model (LEM) of a DE membrane subjected to a static voltage and time-harmonic, uniform pressure loading experienced in acoustic applications. In an effort to adjust the in-plane stress to tune the stiffness of the DE membrane, this paper develops an electroacoustic lumped element model for a uniformly biaxial tensioned DE membrane subject to constant voltage loading for acoustic applications. The resulting model is experimentally validated and thus enables the design of a DE membrane in an acoustic liner application.

The paper is organized as follows. In Section \ref{LEM_sec}, the LEM parameters and fundamental frequency of a DE membrane are derived based on its quasi-static response to a uniform pressure loading. The predicted response is a function of prestretch and applied pressure. Section \ref{Experimental Setup} describes the fabrication procedure of the DE membrane sample as well as the experimental setup using an acoustic impedance tube and laser Doppler vibrometer. The stress dependence on voltage (Eq. \ref{DE stress}) will then be substituted into the resonance expression and compared to experimental results in Section \ref{Results and Discussion} to validate the DE membrane model. Finally, Section \ref{Conclusions} will provide conclusions and future work on this topic. 

\section{Lumped Element Model} \label{LEM_sec}
\begin{figure}[hbt!]
	\centering
	\includegraphics[scale = 0.55]{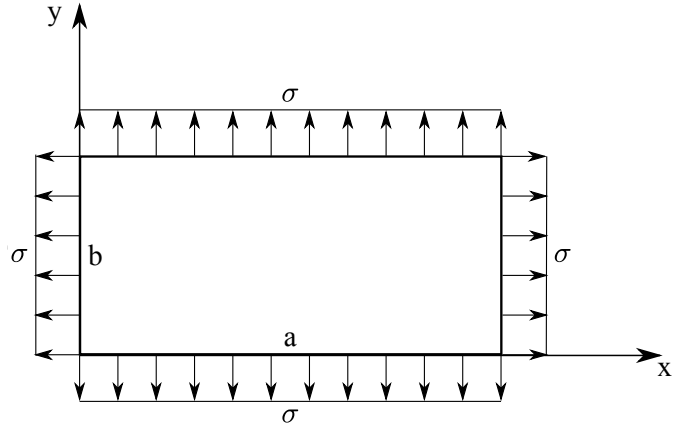}
	\caption{Schematic and coordinate system of a rectangular tensioned DE membrane.}
	\label{Membrane sketch}
\end{figure}

Dielectric elastomers are polymer films with a thickness and bending stiffness dependent on the material. Generally, a DE film can be thinner than 100 microns, especially if prestretched as is the case here. If the material is very thin and cannot support a bending moment, it can be modeled as a tensioned membrane. Therefore, the problem of interest is the 2-D damped wave equation with the configuration as illustrated in Figure \ref{Membrane sketch}.

The governing equation can be written as \cite{leissa2011vibrations}
\begin{equation}
c^2 \nabla^2 w - \frac{\partial^2w}{\partial t^2} - \frac{R}{\rho_a}\frac{\partial w}{\partial t} = \frac{q}{\rho_a},
\label{gov eqn}
\end{equation}
where $c$ is the wave speed of the membrane given by $c^2 = \sigma/\rho$, the stress ($\sigma$) is given by Eq. \ref{DE stress}, $\rho$ is the membrane density, $\rho_a=\rho h$ is the areal density, $w$ is the membrane displacement, $R$ is damping, and $q$ is the applied pressure. The initial stress in Eq. \ref{DE stress} can be expressed by the Yeoh model \cite{holzapfel2000nonlinear} as 
\begin{equation}
\sigma_0(\lambda) = 2 \left( \lambda^2 - \frac{1}{\lambda^4} \right) \sum_{i=1}^3 iC_{i0} \left(I_1 - 3 \right)^{i-1},
\label{Yeoh}
\end{equation}
where the first strain invariant is
\begin{equation}
I_1 = 2\lambda^2 + \frac{1}{\lambda^4},
\end{equation}
and the constants ($C_{i0}$) for Elastosil Film 2030, the material used here, are shown in Table \ref{Elastosil Constants}.
\begin{table}[!htpb]
	\centering
	\caption{Elastosil Film 2030 material constants used in the Yeoh model \cite{hoffstadt2018characterization}.}
	\begin{tabular}{lc} 
		\hline 
		Parameter & Value [kPa] \\
		\hline
		\hline
		$C_{10}$ & 180.7     \\
		$C_{20}$ & -16.7     \\
		$C_{30}$ & 6.6      \\ 	
		\hline
	\end{tabular}
	\label{Elastosil Constants}
\end{table}

The pinned boundary conditions are 
\begin{equation}
w(0,y,t) = w(a,y,t) = w(x,0,t) = w(x,b,t) = 0,
\label{BCs pinned}
\end{equation}
where $a$ and $b$ are the rectangular membrane side lengths as shown in Figure \ref{Membrane sketch}. Assuming a uniform time-harmonic pressure loading, $q_0e^{j\Omega t}$  ($j=\sqrt{-1}$), and using separation of variables, the modal solution is of the form
\begin{equation}
w(\hat{x},\hat{y},\hat{t}) = \sum_{m=1}^{\infty} \sum_{n=1}^{\infty} \sin(m\pi k \hat{x})\sin(n\pi \hat{y}) {T}_{mn}(\hat{t}),
\label{basic sol 1}
\end{equation}
where $\hat{x} = x/b$, $\hat{y} = y/b$, $\hat{t}=\omega_{11} t$, $k=b/a$ is the membrane aspect ratio ($1$ for a square membrane), $q_0$ is a uniform pressure loading, and ${T}_{mn}$ is the time dependent solution. ${T}_{mn}$ may be transformed into a dimensionless displacement given by
\begin{equation}
\hat{T} = \frac{T_{mn} \sigma h }{q_0b^2}.
\label{T hat}
\end{equation}
 Substituting Eq. \ref{basic sol 1} into Eq. \ref{gov eqn} and simplifying yields the dimensionless differential equation
\begin{equation}
\ddot{\hat{T}} + 2\zeta_{mn}\hat{\omega}_{mn} \dot{\hat{T}} + \hat{\omega}_{mn}^2\hat{T} = \hat{q}_{mn}e^{j\hat{\Omega}\hat{t}},\ \ \ m = n = 1,3,5,\dots
\label{time dep}
\end{equation}
The damping ratio ($\zeta_{mn}$), normalized resonance frequency ($\hat{\omega}_{mn}$), and normalized amplitude of the forcing function ($\hat{q}_{mn}$) are
\begin{equation}
\zeta_{mn} = \zeta_{11} \frac{\sqrt{k^2+1}}{\sqrt{m^2k^2+n^2}},
\end{equation}
\begin{equation}
\hat{\omega}_{mn} = \frac{\omega_{mn}}{\omega_{11}} = \frac{\sqrt{m^2k^2+n^2}}{\sqrt{k^2+1}},
\end{equation}
and
\begin{equation}
\hat{q}_{mn} = \frac{16}{mn\pi^4[k^2+1]}.
\end{equation}
Here, $\zeta_{11}$ is the damping ratio for the fundamental mode, $\omega_{mn}$ is the radian resonance frequency that is dependent on mode ($m,n$), $\omega_{11}$ is the fundamental resonance frequency, and $\hat{\Omega} = \Omega/\omega_{11}$. The governing equation is an SDOF system such that the modal resonance frequencies are
\begin{equation}
f_{mn} =  \frac{\omega_{mn}}{2\pi} = \frac{c}{2 b}\sqrt{(m k)^2+n^2}.
\end{equation}
Since $c^2 = \sigma/\rho$, the fundamental resonance frequency, $m=n=1$, for a square membrane ($k=1$) is thus
\begin{equation}
f_{11}=\frac{1}{b\sqrt{2}}\sqrt{\frac{\sigma}{\rho}} \approx \frac{0.707}{b}\sqrt{\frac{\sigma}{\rho}}.
\label{f_11}
\end{equation}
For a uniform pressure loading, $q_0$, the dimensionless static deflection is
\begin{equation}
\hat{\delta}(\hat{x},\hat{y}) = \frac{16}{\pi^4} \sum_{m=1}^{\infty}\sum_{n=1}^{\infty} \frac{\sin(m\pi k \hat{x}) \sin(n \pi \hat{y}) }{mn[(mk)^2+n^2]},\ \ \ m = n = 1,3,5,\dots
\label{static response}
\end{equation}
where $\hat{\delta} = {\delta (\sigma h }/{q_0b^2})$ is the dimensionless deflection.

\subsection{Acoustic Impedance}
The static displacement solution, Eq. \ref{static response}, can be used to determine the acoustic impedance of the DE in terms of lumped parameters as described in Merhaut \cite{merhaut1981theory} and Beranek et al. \cite{beranek2012acoustics}. The lumped acoustic compliance can be calculated by relating it to the ratio of volume displacement to the applied uniform pressure load with no applied voltage
\begin{equation}
C_{aM} = \frac{\Delta \textrm{Vol}}{q_0},
\label{lumped compliance relation}
\end{equation}
where the subscript $a$ refers to acoustic, $M$ refers to membrane, and the variable $\Delta \textrm{Vol}$ is given by
\begin{equation}
\Delta \textrm{Vol} = \frac{q_0b^4}{\sigma h}\int_0^1 \int_0^{1/k} \hat{\delta} d\hat{x}d\hat{y}.
\label{volume velocity}
\end{equation}
Substituting the volume displacement expression into Eq. \ref{lumped compliance relation}, evaluating the summation for $\hat{\delta}$ (Eq. \ref{static response}), and assuming a square membrane ($k=1$) yields
\begin{equation}
C_{aM}  = \frac{0.0351 b^4}{\sigma h}.
\label{lumped compliance}
\end{equation}
The lumped acoustic mass can be found by equating the distributed kinetic energy to that of an acoustic mass, resulting in
\begin{equation}
 M_{aM} = \frac{\rho h b^2}{(\Delta Vol)^2}\int_0^1\int_0^{1/k} \left(\frac{q_0 b^2}{\sigma h}\right)^2 \hat{\delta}^2 d\hat{x}d\hat{y}.
 \label{lumped mass equality}
\end{equation}
Evaluating the integral and simplifying yields the acoustic lumped mass
\begin{equation}
M_{aM} = \frac{1.3785 \rho h}{b^2}.
\end{equation}
The lumped resistance for the fundamental frequency $(m,n)=(1,1)$ is given by
\begin{equation}
R_{aM} = 2\zeta_{11} \sqrt{\frac{M_{aM}}{C_{aM}}}.
\end{equation}
The final acoustic impedance expression is given by
\begin{equation}
Z_{aM} = sM_{aM} + \frac{1}{sC_{aM}} + R_{aM},
\label{Impedance}
\end{equation}
where $s = j \omega$ and $\omega$ is the radian frequency. 
Finally, the resonance frequency in the lumped approximation is given by
\begin{equation}
f_{res} = \frac{1}{2 \pi \sqrt{M_{am}C_{aM}}}=\frac{0.7235}{b} \sqrt{\frac{\sigma}{\rho}},
\label{resonance_LEM}
\end{equation}
which differs from the exact value given in Eq. \ref{f_11} by 2.3\%.

\subsection{Acoustic Radiation Mass}
The physical parameters derived thus far are for a membrane vibrating in a vacuum, while the sample is actually vibrating in air. The air surrounding the membrane exerts a pressure force on it that can be considered a complex radiation impedance, which reduces to an acoustic mass at low values of $\nu b$, where $\nu=\omega/c$ is the acoustic wavenumber. This mass must be included in Eq. \ref{Impedance}. We make the standard assumption here that the radiation mass of a membrane is approximately the same as an infinite baffled piston. The normalized specific acoustic impedance of a rectangular piston at low values of $\nu b$ is given in Mellow et al. \cite{Mellow_2016}. Using this approach, a lumped acoustic radiation mass is given by
\begin{eqnarray}
M_{aP} = 0.946 b \frac{\rho}{2S},
\label{PistonMass}
\end{eqnarray}
where $S=b^2$ is the area of the square membrane. Note that this is for the case of a square piston in an infinite baffle. However, the tested sample in the current study acts as a recessed square piston, where air in the recess is also moving in unison with the piston at low frequencies. The mechanical mass in the recess is $m_{rec} = \rho S t_{rec}$, where $t_{rec}$ is the thickness of the recess. This can be converted to acoustic mass via division by $S^2$. Reorganizing to maintain the same form as Eq. \ref{PistonMass} yields
\begin{equation}
M_{aRec} = \frac{t_{rec}}{b/2} b \frac{\rho}{2S}.
\end{equation}
The total acoustic radiation mass is therefore
\begin{equation}
M_{aRad} = M_{aP} + M_{aRec} = \left( 0.946 + \frac{t_{rec}}{b/2} \right)  \frac{b\rho}{2S}.
\end{equation}
Substituting the recess depth (6.86 mm) and the sample side length (see Table \ref{Membrane dimensions} in the following section) into the parenthesis simplifies the expression to
\begin{equation}
M_{aRad} = 1.486  \frac{b\rho}{2S}.
\end{equation}

\subsection{Voltage Effect}
\begin{figure}[hbt!]
	\centering
	\begin{subfigure}[h]{0.24\textwidth}
		\centering
		\includegraphics[width=\textwidth]{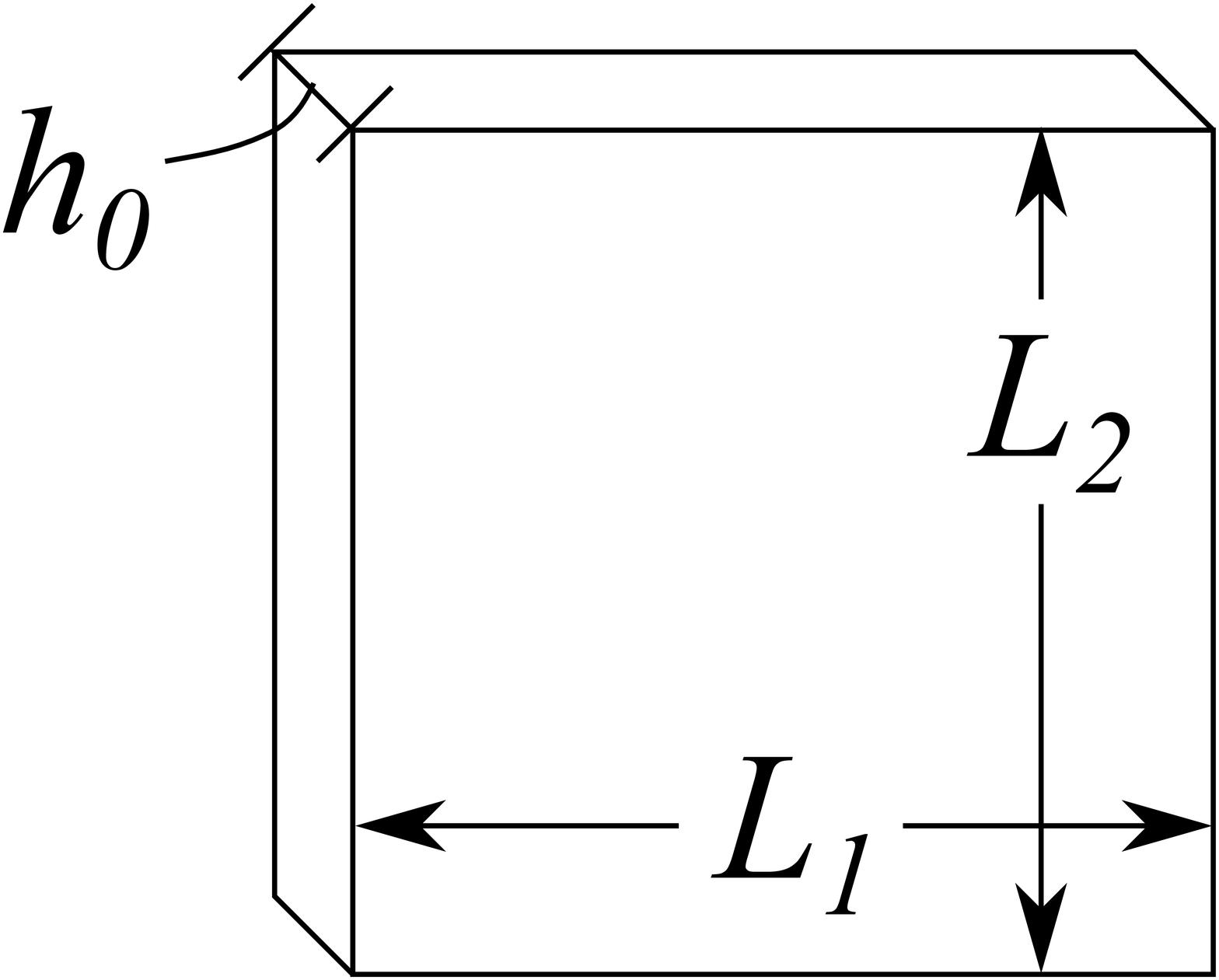}
		\caption{Reference state}
		\label{fig1a}
	\end{subfigure}
	\hfill
	\begin{subfigure}[h]{0.24\textwidth}
		\centering
		\includegraphics[width=\textwidth]{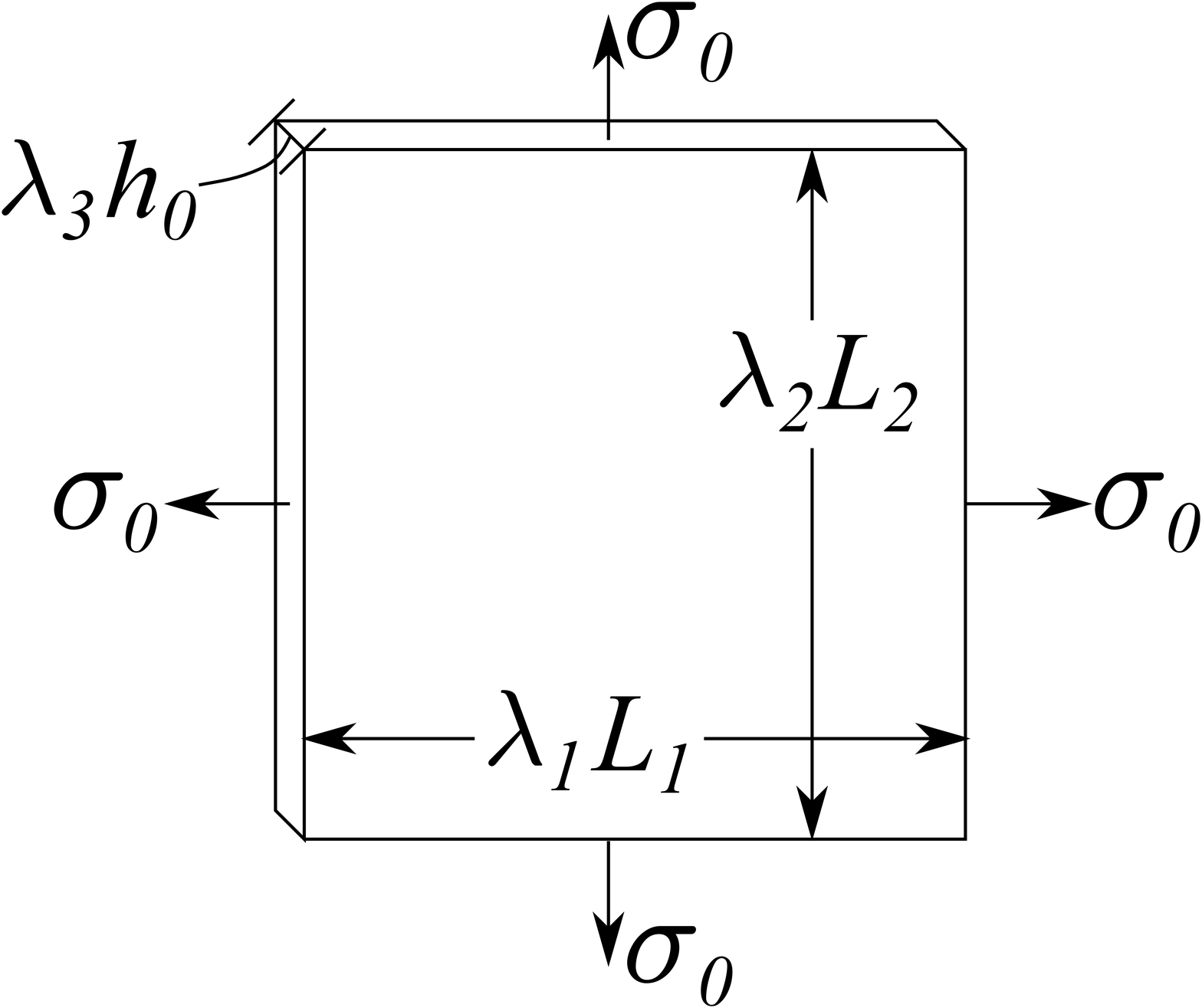}
		\caption{prestretched}
		\label{fig1b}
	\end{subfigure}
\hfill
\begin{subfigure}[h]{0.24\textwidth}
	\centering
	\includegraphics[width=\textwidth]{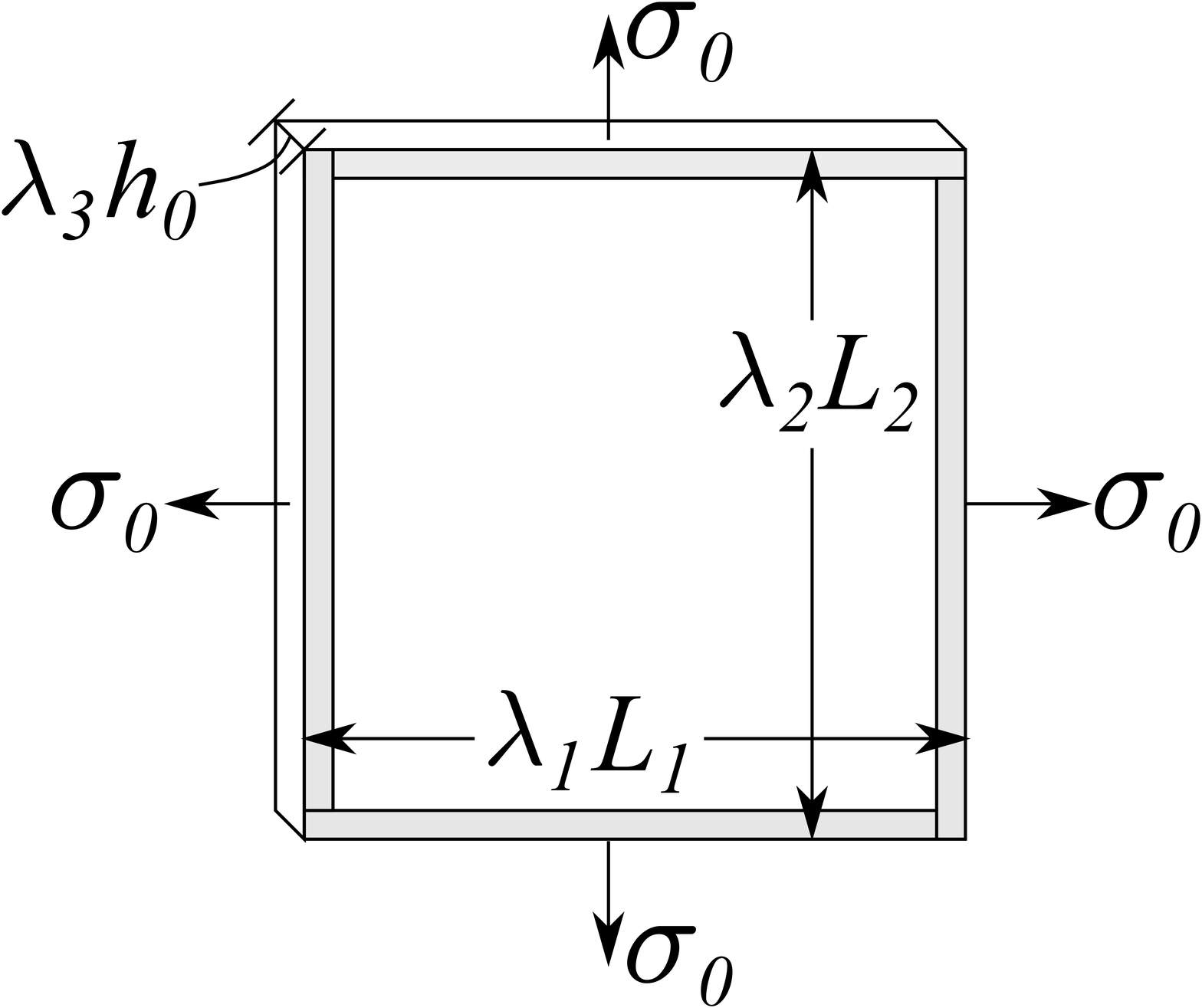}
	\caption{Constrained}
	\label{fig1c}
\end{subfigure}
\begin{subfigure}[h]{0.24\textwidth}
	\centering
	\includegraphics[width=\textwidth]{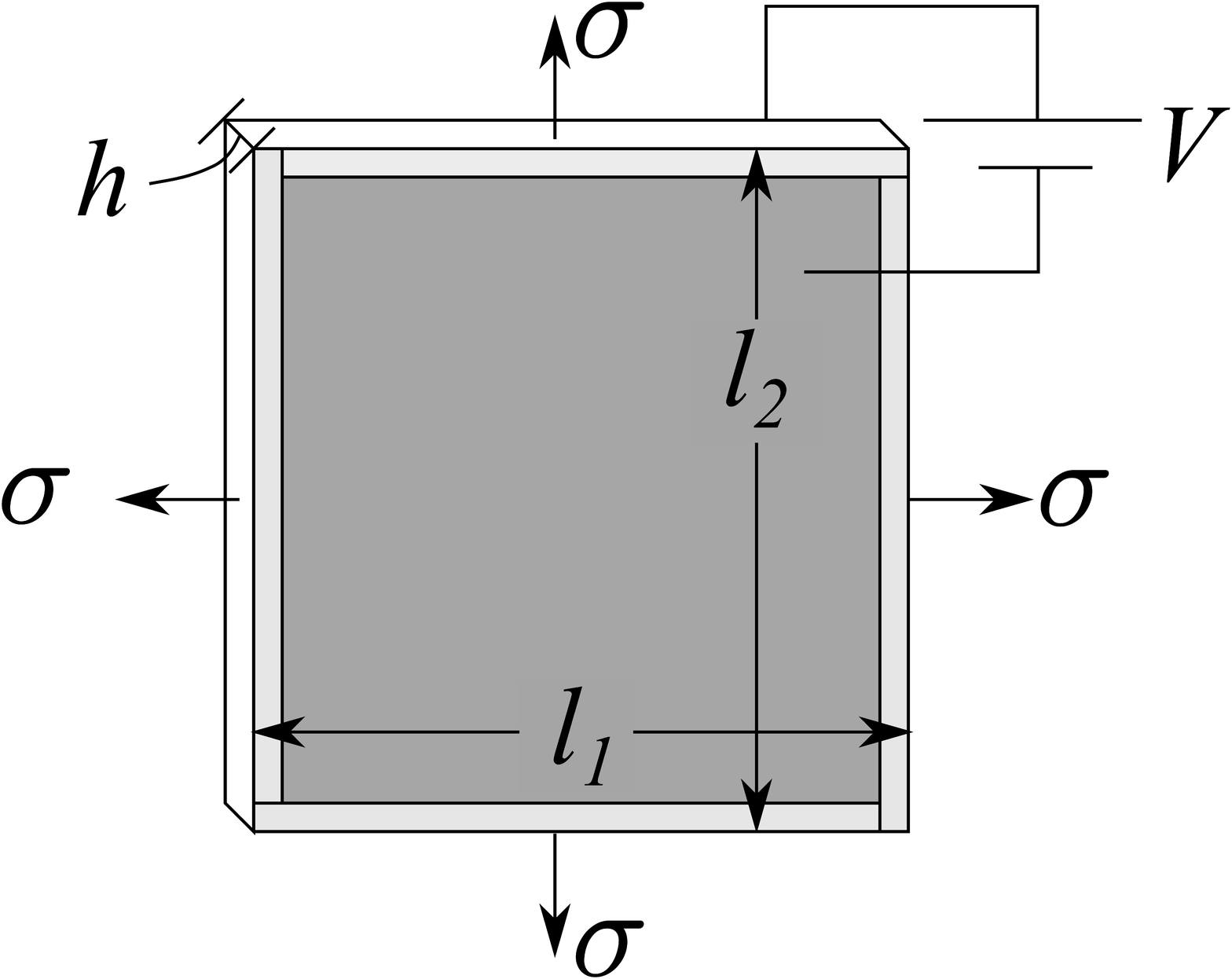}
	\caption{Actuated}
	\label{fig1d}
\end{subfigure}
	\caption{Schematic of the prestretch and voltage actuation process for a DE membrane. (a) In the reference state, the membrane is not stretched. (b) The membrane is prestretched in both directions resulting in an initial in-plane stress of $\sigma_0$. (c) A support frame is added. (c) Grease electrodes are added, and the membrane is subject to a voltage $V$ across its thickness.}
	\label{DE Mechanism}
\end{figure}

The physical mechanism of an actuated DE is  that of a material being squeezed when an electric field is applied across a thin tensioned membrane. This process is analogous to an applied pressure and results in an areal expansion when the material is unconstrained. This results in a reduction of the in-plane stress. An alternative approach is to constrain the DE at its peripheral boundaries as shown in Figure \ref{DE Mechanism}; this approach is adopted in this paper. The DE in its initial (reference) state is shown in Figure \ref{fig1a}. The DE can then be stretched in each direction by some desired prestretch $\lambda_i = l_i/L_i$, where $L$ and $l$ are the before and after stretch dimensions, respectively, and subscript $i$ is the direction index (Figure \ref{fig1b}). The DE is then pinned at its boundaries by a rigid frame as shown in Figure \ref{fig1c}. Finally, a thin, compliant electrode is applied on either side of the DE membrane (Figure \ref{fig1d}) and an electric field is applied across its thickness.

The electrostatic pressure imposed on the DE is \cite{Pelrine_1998}
\begin{equation}
P = \epsilon_0\epsilon_r \Bigg(\frac{V}{h} \Bigg)^2,
\label{DE stress - Pelrine}
\end{equation}
where $\epsilon_0$ is the permittivity of free space, $\epsilon_r$ is the relative permittivity of the material, $V$ is the applied voltage across the membrane, and $h$ is the thickness of the membrane. Equation \ref{DE stress - Pelrine} represents the stress reduction of the initial prestretch imposed on the DE \cite{Suo_2010}. For the pinned configuration shown in Figure \ref{fig1d}, the in-plane stress reduction due to the application of an electric field for an incompressible material is \cite{Lu_2012} 

\begin{equation}
\sigma = \sigma_0 - \epsilon_0 \epsilon_r \Bigg(\frac{\lambda^2}{h_0}V \Bigg)^2,
\label{DE stress}
\end{equation}
where $\sigma_0$ is the initial prestress, $\lambda = \lambda_1 = \lambda_2$ for biaxial prestretch, and $h_0$ is the initial thickness of the unstretched DE membrane.

The effect of voltage can be incorporated into the membrane impedance. Neglecting the mass of the electrodes, the applied voltage only modifies the membrane stress. Hence, only the lumped compliance term needs to be updated. This is accomplished by noting that $h=h_0/\lambda^2$ and substituting Eq. \ref{DE stress} into the lumped compliance expression, Eq. \ref{lumped compliance},
\begin{equation}
C_{aM}  = \frac{0.0351 \; b^4 \; \lambda^2/h_0}{\sigma_0 - \epsilon_r \epsilon_0 \Bigg(\frac{\lambda^2}{h_0}V \Bigg)^2}.
\label{lumped compliance - v2}
\end{equation}

The resonance frequency expression can be updated accordingly. The fundamental resonance frequency of the membrane in Eq. \ref{resonance_LEM}, can be rewritten as
\begin{equation}
f = \frac{1}{2\pi}\sqrt{\frac{1}{C_{aM}(M_{aM}+M_{aRad})}}.
\label{Res no E}
\end{equation} 
Substituting in Eq. \ref{lumped compliance - v2} yields
\begin{equation}
f = \frac{1}{2\pi} \sqrt{\frac{\sigma_0 - \epsilon_r\epsilon_0 \left(\frac{\lambda^2}{h_0}V \right)^2} {0.0351 \; b^4 \; \lambda^2/h_0 \; (M_{aM}+M_{aRad}) }},
\end{equation}
and normalizing by the unactuated ($V=0$) resonance expression yields
\begin{equation}
\frac{f(V)}{f(V=0)} = \sqrt{\frac{{\sigma_0 - \epsilon_r\epsilon_0 \left(\frac{\lambda^2}{h_0}V \right)^2}}{\sigma_0}}.
\label{Res E}
\end{equation}
The relative permittivity ($\epsilon_r$) is a function of both material properties and prestretch. For the material used here, Elastosil Film 2030, the relative permittivity was given in  Hodgins et al. \cite{Hodgins_2016} as a function of the prestretch
\begin{equation}
\epsilon_r = -0.28\lambda + 2.76
\label{relative permittivity expression}
\end{equation}
This allows us to compare the resonance frequency variation of DE membranes as a function of both prestretch and voltage.

\section{Experimental Setup} \label{Experimental Setup}

\subsection{Sample Preparation} \label{Sample Preparation}
\begin{figure}[hbt!]

	\centering
	\includegraphics[scale = 0.65]{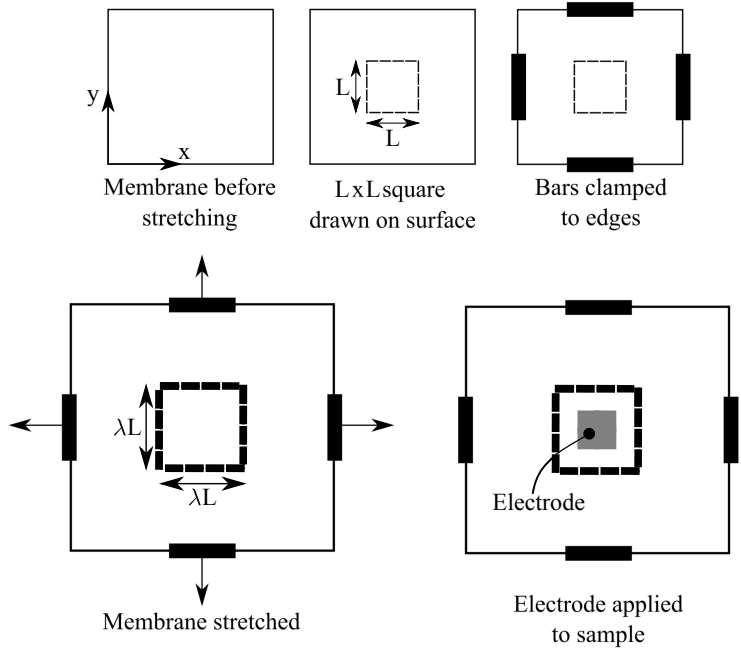}
  \vspace{-0.2in}
	\caption{Steps for preparing a prestretched square DE membrane.}
	\label{pre-stretch}
  \vspace{-0.2in}
\end{figure}

The sample is a 25.4 mm square DE membrane. The material used is Elastosil Film 2030, which has an initial-unstretched thickness, $h_0$, of 100 microns. Before the DE is sandwiched between the two square frames, it is prestretched by the same factor, $\lambda$, in both directions. Figure \ref{pre-stretch} 
shows the steps: 1) the DE membrane is laid on a flat surface; 2) an $L \times L$ square is sketched onto the surface before 3) clamping the DE on its sides and stretching. The stretching is done in two parts: the edges along direction $x$ are stretched first followed by direction $y$ to minimize tearing. While stretching, the original marker lines drawn on the sample (the $L \times L$ square) are stretched as well. This results in thicker reference lines, which makes it difficult to accurately estimate the stretched dimensions. This can be seen in Figure \ref{pre-stretch} as the exaggerated difference in the thickness of the dashed lines. The maximum uncertainty of the prestretch is approximately $\pm0.15$, caused by enlargement of the marker lines. Finally, when the nominal desired stretch is reached, the electrode is applied if needed. The DE is then adhered to the rigid square frame using Very High Bond (VHB) tape.

Two types of samples are made, one with electrodes and one without. The type without electrodes has a dot(s) added to its surface using a thin silver sharpie to reflect light back to the laser vibrometer. The other sample is coated with a thin layer of carbon grease electrode. The coating is applied using a cotton swab and excess is removed with a soft spatula, resulting in a very thin layer. Carbon grease is commonly used by researchers \cite{Pelrine_1998, wissler2007modeling} and is readily available for purchase. Table \ref{Membrane dimensions} provides dimensions of the sample and density of both the membrane and electrodes. Note that the density of the DE membrane and carbon-grease electrode are very close. Finally, copper tape is used to connect the edge of the grease electrodes to the wires of a high voltage amplifier (Trek Model 10/40A). 
\begin{table}[!htpb]
	\centering
	\caption{Material properties and dimensions of DE membrane sample.}
	\begin{tabular}{l|lll} 
		\hline 
		& Description & Variable & Value [unit] \\
		\hline
		\hline
		\multirow{3}{6em}{DE: Elastosil Film 2030} & side length                  & $a, b$     & 25.4 [mm]     \\
		& thickness (prior to stretch)    & $h_0$      & 100  [$\mathrm{\mu}$m] \\
		& density                         & $\rho$     & 1000 [kg/m$^3$] \\ 
		\hline
		Electrodes & density    & $\rho_E$        & 1010 [kg/m$^3$]      \\
		\hline
		Frame & side length & $a, b$     & 25.4 [mm]     \\
		 & thickness & $t_{rec}$ & 6.86 [mm] \\
		\hline
	\end{tabular}
	\label{Membrane dimensions}
\end{table}

\subsection{Acoustic Tube and Vibrometer Setup}
\begin{figure}[hbt!]
	\centering
	\includegraphics[scale = .8]{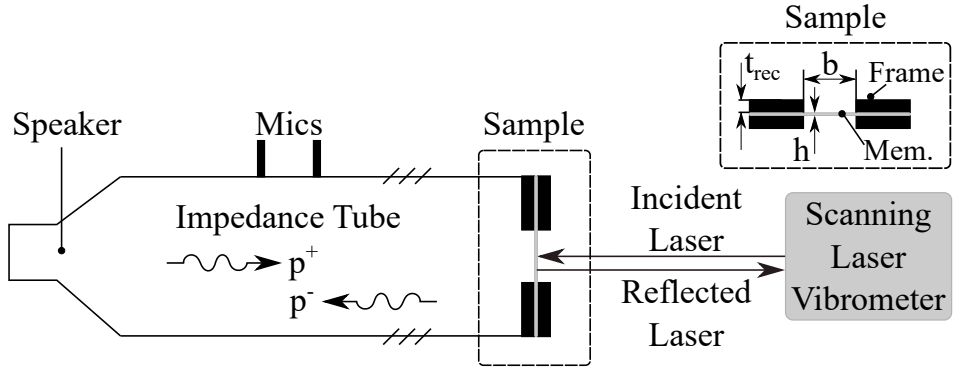} 
	\caption{Sketch of the normal incidence tube setup with a vibrometer setup behind to measure DE vibrations.}
	\label{Setup}
\end{figure}

A schematic of the experimental setup is shown in Figure \ref{Setup} and depicts the Two Microphone Method (TMM) to determine the acoustic impedance of the sample \cite{astm1998standard}.
The Brüel and Kjær (B\&K) wide-spacing large circular impedance tube type 4206 with a diameter of 100 mm is used. The length between the catenoidal horn exit and the test specimen is 875 mm. Microphone 1 is closest to the speaker and the second microphone is 100 mm away from the first microphone, which is 575 mm from the test specimen. The signal sent to the speaker is either a single-frequency sinusoid or psuedo-random periodic noise covering an octave band, where the SPL at the sample face is held constant for each frequency. 

Normal velocity measurements of the sample are taken along the center line every 1.25 mm using a Polytec scanning laser vibrometer, type PSV 300. A laser Doppler vibrometer measures the frequency shift of a laser by the Doppler effect, which can be used to find the velocity at a point. The measurement is synced with the pseudo-random periodic noise signal sent to the speaker using a trigger signal. The standoff distance between the DE membrane sample and vibrometer is between 21.6 cm and 55.3 cm. The data is sampled at 16384 Hz and analyzed using a DFT with a frequency resolution of 1 Hz. Forty spectral averages are taken to compute the frequency response between the speaker excitation and the velocity response at multiple points.


\section{Results and Discussion} \label{Results and Discussion}
The LEM prediction of the displacement response of the sample is compared to experimental measurements in the following section. The first step is to determine the prestretch applied to the sample. This is then substituted in the model of the static response (Eq. \ref{static response}) and resonance frequency (Eq. \ref{Res E}) of the membrane, from which the effect of the excitation voltage can be determined. 

\subsection{Membrane prestretch}
Although an approximate value of prestretch is known using the sketch drawn on the membrane surface, the actual prestretch falls within a range of values as explained in Section \ref{Sample Preparation}. The precise prestretch value is therefore estimated by minimizing the differences between the computed static displacement and resonance frequency, and the experimentally observed values. Both quantities are directly related to the in-plane stress (Eqs. \ref{static response} and \ref{resonance_LEM}, respectively), which is a function of prestretch via Eq. \ref{Yeoh}. An example computation of the displacement and resonance frequency is shown here.

\begin{figure}[hbt!]
	\centering
	
	\includegraphics[scale =0.7]{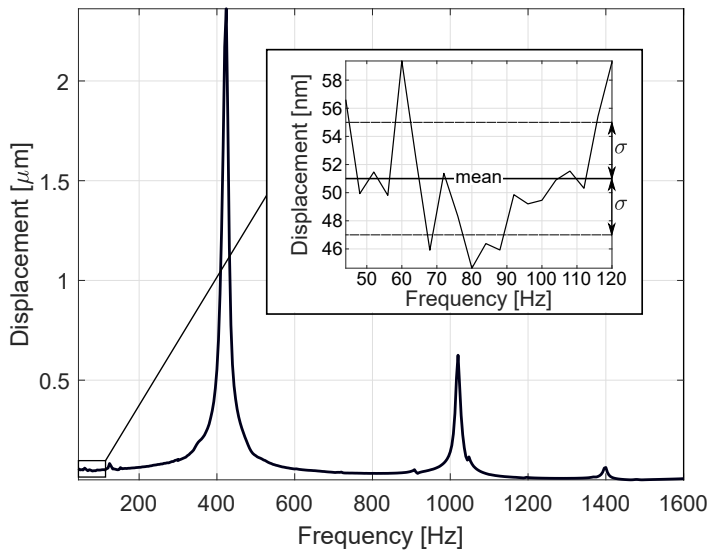}
	\caption{Measured displacement frequency response at the center of the membrane. Mean = 51 nm; standard deviation $\sigma$ = 4 nm.}
	\label{Frequency Response}
\end{figure}

The frequency response of the membrane center for an 80 dB pressure loading is plotted in Fig. \ref{Frequency Response}. The static or dc response is estimated by measuring the center displacement at low frequencies where the response asymptotes to a constant. A zoomed view of the frequency response below 120 Hz where the response asymptotes is shown in Fig. \ref{Frequency Response}. The mean of the response in this frequency range is approximately 51 nm with a standard deviation of 4 nm. Comparison to the static solution theory (Eq. \ref{static response}) will be for a selected frequency in this range, 100 Hz for the analysis here. Figure \ref{Frequency Response} also shows three peaks in the response. The first peak, and the global maximum, is the fundamental frequency used for comparison to the LEM resonant frequency Eq. \ref{Res no E}. The two smaller peaks are higher-order modes where the center response reaches a relative maximum. 

\begin{figure}[hbt!]
	\centering
	\includegraphics[scale = 0.7]{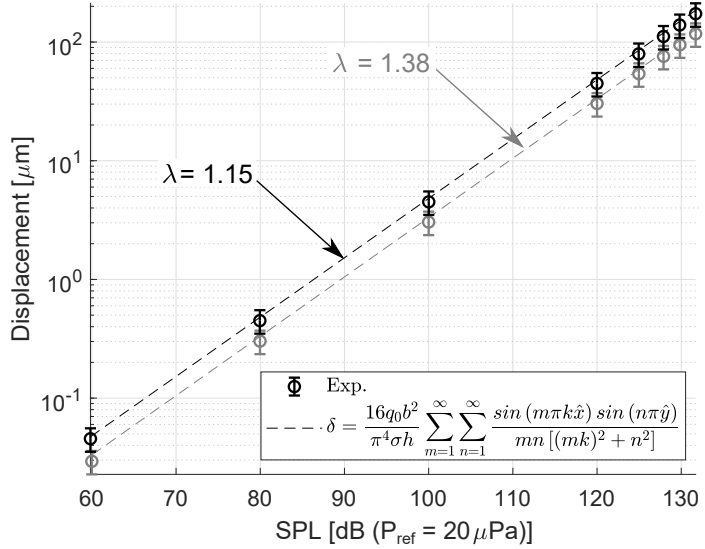}
	\caption{Measured membrane displacement response at 100 Hz compared to the theoretical static membrane response for two prestretch values.}
	\label{Center response - linearity}
\end{figure}

The center response is measured for two prestretches at 100 Hz for eight SPLs and is plotted in Fig. \ref{Center response - linearity}.
Each of the eight responses is compared to the static displacement solution and the prestretch is estimated by setting Eq. \ref{static response} equal to the measured response, resulting in eight prestretch values. The average prestretch is computed and shown in the second column of Table \ref{pre-stretch Estimation}.  For a visual prestretch of 1.1 and 1.3, the approximate prestretch using the static displacement solution is 1.17 and 1.46, respectively.

\begin{table}[hbt!]
	\centering
	\caption{Prestretch estimated from visual inspection, static displacement, resonance frequency, and the difference minimization considering both static displacement and resonance frequency.}
	\begin{tabular}{r|ccc} 
		\hline 
		Visual & Static Disp. & Resonance & Difference Min. \\
		\hline
		\hline
		1.1 & 1.17 & 1.13 & 1.15 \\
		\hline
		1.3 & 1.46 & 1.29 & 1.38 \\
		\hline
	\end{tabular}
	\label{pre-stretch Estimation}
\end{table}

The frequency response is also measured and the resonant frequency identified as 390 Hz and 530 Hz for a visual prestretch of 1.1 and 1.3, respectively. The frequency value is accurate to $\pm$ 0.5 Hz for the 1 Hz frequency resolution of the measurement. The prestretch is numerically determined by setting the measured resonance equal to Eq. \ref{Res no E}, noting the negligible difference between the natural and resonance frequency due to very low damping. For the visual prestretches in Table \ref{pre-stretch Estimation}, the approximate prestretch corresponding to the measured resonance frequency is 1.13 and 1.29, respectively. These values are lower than those from the static solution and closer to the visual prestretch values.

To minimize the overall uncertainty, the prestretch is varied over a range of values and the relative-difference in both the static deflection 
and the measured resonance frequency is calculated. The total difference of the two is taken as the square root of the sum of the squared differences and the local minimum is found. The results are listed in the last column of Table \ref{pre-stretch Estimation} and happen to be the averaged values between the two methods because the slope near the intercept is linear. The updated prestretch values are approximately 1.15 and 1.38 for the visual prestretches of 1.1 and 1.3, respectively. The updated prestretch values will be called ``prestretch'' for the remainder of the paper. The static displacement and resonant frequency predictions will now be updated with the prestretch values for comparison to the experimental results.

The experimentally measured center response at 100 Hz is plotted versus incident SPL together with the model (Eq. \ref{static response}) in Fig. \ref{Center response - linearity}.
The experimental values are plotted with random uncertainty estimates with a 95\% confidence level calculated based on the procedures in Bendat et al. \cite{bendat2011random}. The model shape aligns with the measured response and falls within the estimated uncertainty, indicating an accurate prediction of the prestretch and static displacement amplitude. The membrane response remains linear, tracking the modeled response up to at least a 132 dB pressure loading, the maximum achievable SPL. Additionally, the resonant frequency is calculated based on the estimated prestretch value with the results shown in Table \ref{Resonance}.
The resonance is over predicted by a maximum of 9.4\% for the cases tested, which is deemed reasonable for a lumped element model.
\begin{table}[hbt!]
	\centering
	\caption{Resonance frequency measured vs. calculated using the minimized composite difference value from column four of Table \ref{pre-stretch Estimation}.}
	\begin{tabular}{l|lll} 
		\hline 
		Prestretch & Measured & Calculated & Difference \\
		\hline
		\hline
		1.15 & 390 & 417 & 6.9\% \\
		\hline
		1.38 & 530 & 580 & 9.4\% \\
		\hline
	\end{tabular}
	\label{Resonance}
\end{table} 

\subsection{Model Validation}
\begin{figure}[hbtp!]
	\centering
	\vspace{-.1in}
	\includegraphics[scale = 0.45]{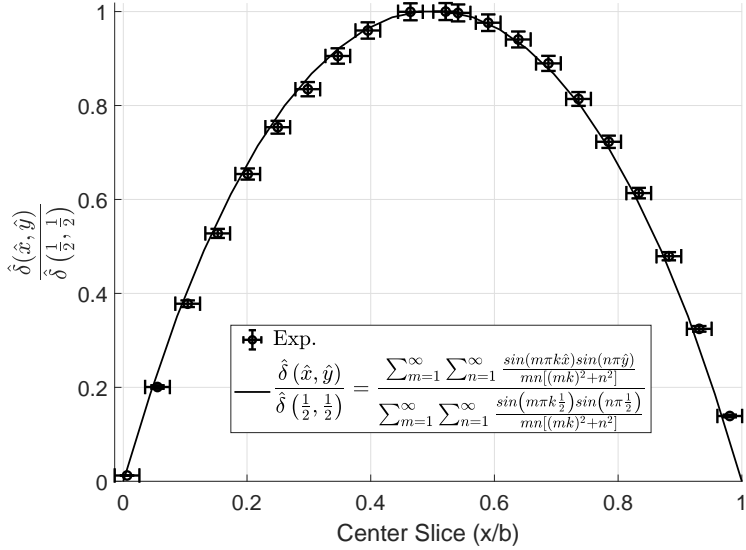}
	\caption{Normalized displacement of the measured membrane response at 100 Hz and theoretical static membrane response for a prestretch of 1.38.}
	\label{Centerline response}
\end{figure}

The center line displacement of the DE membrane is measured for validation of the static mode shape. The laser vibrometer measures the response across the surface at a sinusoidal forcing of 100 Hz and an incident pressure of 80 dB.
Figure \ref{Centerline response} plots the normalized displacement response and the static model prediction given by Eq. \ref{static response}. The experimental values are plotted with random uncertainty estimates with a 95\% confidence level calculated based on the procedures in Bendat et al. \cite{bendat2011random}. 
Uncertainty in the $x$-axis values is determined via consideration of mapping the scanning points from the camera using the estimated control target locations. Considering the uncertainties in the measurements, the mode shape agrees well with the analytical model predictions. This validates the mode shape predicted by the model while the amplitude itself is validated in the previous section, as shown in Fig. \ref{Center response - linearity}. 

\subsection{Membrane Actuation}
\subsubsection{Added Mass due to Electrodes}
\begin{figure}[hbt!]
	\centering
	\includegraphics[scale = 0.7]{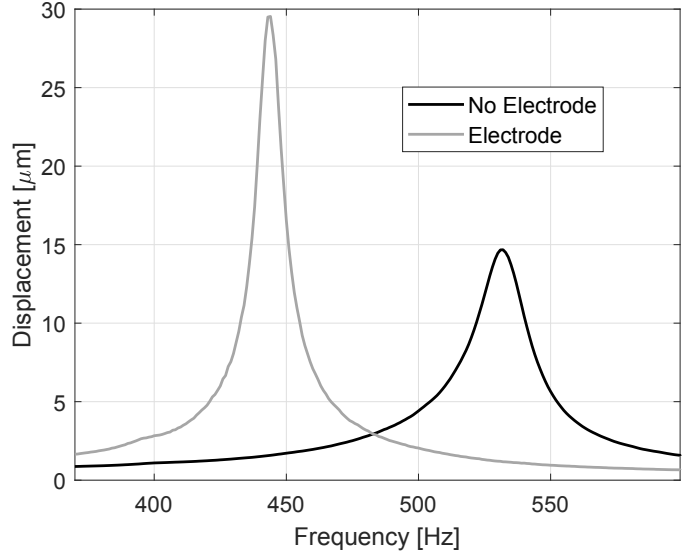}
	\caption{Measured response of the DE membrane with and without carbon grease electrode for a prestretch of 1.38.}
	\label{greaseVnogrease}
\end{figure}

To actuate the DE membrane, electrodes must be added to its surface. This affects the membrane response since the membrane is very thin (less than 100 $\mu$m thick) and has approximately the same density as the electrodes used. To understand how the electrode loads the membrane, the measured frequency response of the membrane with and without the carbon grease electrodes is plotted in Figure \ref{greaseVnogrease}. It was assumed that the electrodes only modified the mass and not the stiffness of the membrane. Starting with Eq. \ref{Res no E}, we thus assume that $M_{aM}$ is affected. The resonance frequencies of the membrane with and without electrodes can extracted from Figure \ref{greaseVnogrease} as 444 Hz and 532 Hz, respectively, and are documented in Table \ref{lumped parameters}.

\begin{table}[!htpb]
	\centering
	\caption{Resonance frequency measured in experiments and lumped parameter values.}
	\begin{tabular}{r|lcc} 
		\hline 
		& Description & Variable & Value [units] \\
		\hline
		\hline
		\multirow{3}{4.5em}{DE Membrane} & Resonance w/o Electrodes & $f$ & 532  [Hz] \\
		& Resonance w/ Electrodes & $f_{E}$ & 444 [Hz] \\ 
		& Lumped acoustic mass & $M_{aM}$     &  	109.0 [kg/m$^4$]	   \\
		\hline
		\multirow{2}{4.5em}{Electrodes} & Added acoustic mass& $M_{aE}$ & 78.2 [kg/m$^4$]      \\
		& Thickness & $h_E$ & 25 [$\mu$m] \\
		\hline
	\end{tabular}
	\label{lumped parameters}
\end{table}

The resonance frequency expression (Eq. \ref{Res no E}) can be updated to include the added electrode mass, $M_{aE}$,
\begin{equation}
f_{exp,E} = \frac{1}{2\pi} \sqrt{\frac{1}{C_{aM}\left(M_{aM} + M_{aRad} + M_{aE}\right)}}.
\label{Res air electrodes}
\end{equation}
The ratio of the added acoustic mass to that of the membrane as a function of the observed fractional change, $df/f$, in the resonance frequency can be expressed as
\begin{equation}
\frac{M_{aE}}{M_{aM}+M_{aRad}} = \frac{1}{\left( 1-\frac{df}{f} \right)^2} -1.
\label{added mass}
\end{equation}
The approximate value of the lumped mass due to the electrodes is 78.2 kg/m$^4$ using Eq. \ref{added mass} and is shown in Table \ref{lumped parameters}. As a check, the mechanical mass (in kg) from the electrodes can be estimated by multiplying the added mass by the area squared ($b^4$) and dividing by the electrode density. The thickness of the electrode layer on either side of the DE membrane can then be estimated. Substituting the values from Tables \ref{Membrane dimensions} and \ref{lumped parameters}, the thickness of the electrode layer is approximately 25 $\mu$m. This value is consistent with the thickness of the carbon-grease electrode estimated via visual inspection.

\subsubsection{Actuation}
\begin{figure}[hbt!]
	\centering
	\includegraphics[scale = 0.7]{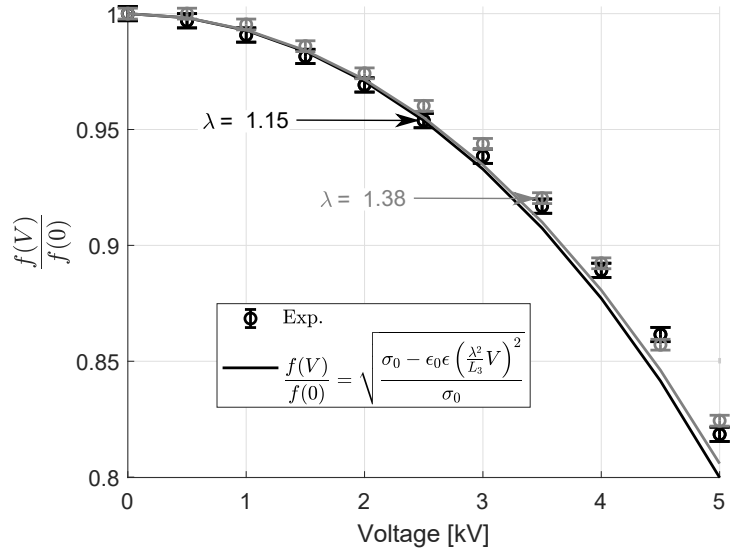}
	\caption{Normalized resonance frequency versus voltage for the experiment and model.}
	\label{Voltage}
\end{figure}

The normalized resonance frequency for both the experiment and the model (Eq. \ref{Res E}) versus voltage is shown in Figure \ref{Voltage}. The resonant frequency is normalized by the no voltage case and so the value at zero voltage is unity. The resonance decreases quadratically with voltage, as expected based on Eq. \ref{Res E}. There is good agreement between experiments and theory for a majority of the range. The maximum difference for $\lambda = 1.15$ and $\lambda = 1.38$ is 2.3\% at 5 kV. This is the same difference as was found between the LEM resonance frequency (Eq. \ref{resonance_LEM}) vs. the exact resonance frequency (Eq. \ref{f_11}). This indicates that the membrane model of the DE is appropriate for the material used here.

\section{Conclusions} \label{Conclusions}

An LEM of a square dielectric elastomer membrane was derived. The damped wave equation in Cartesian coordinates was used to derive the static deflection of a square membrane. Using the relationship between pressure loading and volume displacement as a function of static deflection, the LEM was created. The effect of a dielectric elastomer was added to the model via an actuation term expressed as voltage. Voltage was incorporated into the stress term, which for the LEM affects the membrane's acoustic compliance (i.e., 1/stiffness). The resonant frequency expression was updated accordingly and reduces quadratically with voltage. Material properties and dimensions are parameters that may be chosen to achieve the desired impedance or resonant properties. Then the dimensions, prestretch, and voltage can be designed to achieve a desired impedance or resonance of the DE membrane. 

A scanning laser vibrometer was used to validate the static response of the membrane. The membrane was subjected to various sound pressure levels and behaved linearly over the entire range tested. This validates the model up to 132 dB pressure loading, the maximum achievable pressure in the experimental setup. In order to actuate the DE membrane, electrodes were added to its surface. Therefore, the impedance was modified to include the effects of electrode mass.
The model and experimental resonance frequency had good agreement with a maximum difference of 2.3\% at 5 kV of applied voltage for the two prestretches tested. This validates the use of the membrane approximation for the DE used here.

With the validation of the model, a rectangular DE membrane may now be incorporated as a lumped element of a distributed electroacoustic system. This provides analytic scaling and rapid parametric studies for the initial phase of design. We are interested in using the LEM to predict the impedance of acoustic liners with an active embedded membrane. 
Further development will incorporate an optimization scheme to maximize sound absorption or impedance tuning of the liner, combined with experimental demonstrations of the optimized design(s). 

\section*{References}
\newcommand{\newblock}{}

\end{document}